\documentclass[9.5pt,conference]{IEEEtran}

\usepackage{cite} 
\usepackage{verbatim} 
\usepackage{amsmath,amssymb,amsfonts,amsthm}
\usepackage{algorithmic}
\usepackage{graphicx}
\usepackage{textcomp}
\usepackage[table]{xcolor} 
\usepackage[keeplastbox]{flushend} 

\usepackage{listings} 

\usepackage{cancel} 

\usepackage{url} 

\IEEEoverridecommandlockouts

\newtheorem{theorem}{Theorem}[section]

\usepackage{thmtools}
\declaretheoremstyle[
  spaceabove=\topsep, spacebelow=\topsep,
  headfont=\normalfont\scshape,
  notefont=\mdseries, notebraces={(}{)},
  bodyfont=\normalfont,
  postheadspace=1em,
  qed=\qedsymbol
]{myexample}

\declaretheorem[style=myexample]{example}

\newcommand{\onebar}{\ensuremath{\bar{1}}} 
\newcommand{\UU}{\onebar} 

\DeclareMathOperator{\sgn}{sgn} 
\DeclareMathOperator{\size}{size} 
\DeclareMathOperator{\width}{wd} 
\DeclareMathOperator{\ulp}{ulp} 

\newcommand{\nondegenterngates}{19,\!632} 
\newcommand{\nondegenterngatescmu}{726} 
\newcommand{\nondegenterngatesnoncmu}{18,\!906} 

\usepackage{mathtools} 
\DeclarePairedDelimiter\ceil{\lceil}{\rceil} 
\DeclarePairedDelimiter\floor{\lfloor}{\rfloor} 

\usepackage{accents}

\usepackage{listings}

\usepackage{caption} 

\lstdefinestyle{style_command}
{
  xleftmargin=20pt,
  basicstyle=\sffamily, 
  stepnumber=1, 
  numbersep=10pt,                         
  tabsize=4, 
  extendedchars=true, 
  breaklines=true, 
  captionpos=t, 
  stringstyle=\color{white}\ttfamily, 
  showspaces=false,
  showtabs=false, 
  showstringspaces=false 
}

\lstdefinestyle{style_betweenlines}
{
  xleftmargin=17pt,
  frame=top, 
  frame=bottom,  
  framexleftmargin=17pt,
  framexrightmargin=17pt,
  framexbottommargin=5pt,
  framextopmargin=5pt,
  basicstyle=\sffamily, 
  stepnumber=1, 
  numbersep=10pt,                         
  tabsize=4, 
  extendedchars=true, 
  breaklines=true, 
  captionpos=t, 
  stringstyle=\color{white}\ttfamily, 
  showspaces=false,
  showtabs=false, 
  showstringspaces=false 
 }

\lstdefinestyle{style_betweenlines_header}
{
  xleftmargin=17pt,
  xrightmargin=17pt,
  frame=top, 
  frame=bottom,  
  framexleftmargin=17pt,
  framexrightmargin=17pt,
  framexbottommargin=5pt,
  framextopmargin=5pt,
  basicstyle=\sffamily, 
  stepnumber=1, 
  numbersep=10pt,                         
  tabsize=4, 
  extendedchars=true, 
  breaklines=true, 
  captionpos=t, 
  stringstyle=\color{white}\ttfamily, 
  showspaces=false,
  showtabs=false, 
  showstringspaces=false 
 }

\DeclareCaptionFormat{listing}{\rule{\textwidth}{0.4pt}\par\vskip1pt#1#2\hfill#3}

\lstnewenvironment{command}[1][] {
	\lstset{#1, style=style_command} 
	}{}

\lstnewenvironment{betweenlines}[1][] {
	\lstset{#1, style=style_betweenlines}
	}{}

\lstnewenvironment{betweenlines_header}[1][]
	{\lstset{#1, style=style_betweenlines_header}}
	{}

\begin{document}

\title{
A Tapered Floating Point Extension for the Redundant~Signed~Radix~2~System %
Using the Canonical~Recoding %
}

\author{
\IEEEauthorblockN{Lucius T. Schoenbaum}
\IEEEauthorblockA{\textit{Department of Electrical and Computer Engineering} \\
\textit{University of Alabama in Huntsville}\\
Huntsville, Alabama, 35805-1911 \\
lts0016@uah.edu}
}



\maketitle
\IEEEpubidadjcol



%

\begin{abstract}
A tapered floating point encoding %
is proposed which uses the redundant signed radix~2 system %
and is %
based on the canonical recoding. %
By making use of ternary technology, %
the encoding has a dynamic range exceeding that of %
the IEEE 754-1985 Standard 
for Floating Point Arithmetic (IEEE-754-1985), %
and precision equal to or better than that of  %
the IEEE-754-1985 system %
and the recently proposed Posit system %
when equal input sizes are compared. %
In addition, the encoding is capable of supporting several proposed extensions, %
including extensions to integers, boolean values, complex numbers, %
higher number systems, %
low-dimensional vectors, and system artifacts such as machine instructions. %
A detailed analytic comparison is provided %
between the proposed encoding, %
the IEEE-754-1985 system, %
and the recently proposed Posit number system. %
\end{abstract}


\section{Introduction}\label{s.i}

In recent years, an increase in demand for %
computational numerical processing has been observed. %
In particular, machine learning has become %
widely adopted as a tool for scientific, industrial, and commercial applications, 
including autonous systems. %
This has raised concerns about the energy- and resource-consuming nature of %
the computationally intensive training of neural networks \cite{GreenAI}, %
and the energy and economic cost of data storage, including numerical data, %
by cloud storage service providers \cite{DoEtalStorage, LindstromEtalTradeoff}. %
This has contributed to a recent revival of research into tapered %
floating point number systems \cite{GustYo1, LinstromLloydHitt1, Lindstrom1}. %
The tapered exponent mechanism, %
first proposed by Morris in 1971 \cite{Morris1}, %
and reappearing %
in subsequent proposals by Hamada (URRs) \cite{Hamada83}, %
and Yokoo (extended and variant URRs) \cite{Yokoo92}, %
boosts the precision in %
a range where, presumably, values are most frequent and most likely to occur, 
and provides gradual/graceful underflow and overflow %
while avoiding the exceptionality and inconvenience of a subnormal range. %
This trade-off means that it may be possible to achieve effective training %
of neural networks even after decreasing the input size. %
It is also a convenient property for applications in the growing fields of %
IoT and AI on the Edge \cite{QNN1,CMixNN1}. %


The Posit system \cite{GustYo1, PositStandardDocs3-2} is a tapered system proposed in 2017 as a %
drop-in alternative to the widely adopted IEEE-754-1985 %
Standard for Floating-Point Arithmetic \cite{IEEE754_1985} %
and its 2008 revision \cite{IEEE754_2008}. %
The Posit/Unum system has since %
been studied for its potential in low-precision deep learning applications (AI on the edge) %
\cite{CaLaKhLiGustKu1, LuFaXuLiWa1}, %
for numerical and scientific applications \cite{BuDoShSh1, KlDuPa1}, %
and its hardware costs have been studied \cite{UgFoDi1, LaPaKu1}. %
%
The Posit system can be viewed as a tapered floating point system that %
expresses the most significant part of the exponent in signed base~1, %
rather than base~2. %
This avoids the need to set aside some bits to record the tapered length, %
as was the case in the first tapered system proposed by Morris. %
The Posit system offers enhanced best-case precision and %
potentially enhanced dynamic range relative to %
the fixed-exponent size IEEE-754 system. %
A recent independent study \cite{BuDoShSh1} %
finds that, if inputs are scaled to take advantage of %
the optimal range for the Posit system, %
the results are numerically superior to equally-sized IEEE-754 floats. %

Both the Posit system and the IEEE-754 standard %
are binary specifications, as is the Morris system. %
In this article, we propose a different tapered floating point system %
based on ternary technology. %
We use this technology, the redundant signed radix~2 system, %
and the canonical recoding (also known as the nonadjacent form) %
to provide the basis for a floating point system. %
In its rounding policy and general outlook towards %
numerical exception-handling, the proposed system %
is in close agreement with the recent Posit system, %
as expressed in the work of Gustafson. %

In the report below, we provide data on the precision as a function of input, %
over the entire dynamic range for the Posit system, the IEEE-754 system, %
and the newly proposed system, as well as a range of factors of merit, %
for a side-by-side comparison of the three systems. %
The data indicates that for all input values $x$ in its dynamic range, %
precision after rounding in the system is always (for all $x$) equal to or better than %
the precision of the %
fixed-width binary system (the IEEE-754 system), %
and that of the tapered binary system (the Posit system). %
In the mid-range to approaching the extremities of the total dynamic range, %
use of half-precision in the proposed system can provide higher precision than the %
use of full precision in either of its counterparts. %
For example, within the limits of the dynamic range of the %
IEEE-754 system for 32 bit size (single precision), the proposed system %
with input size 16, which has a larger dynamic range, has a worst-case precision of 8 base-2 digits. %
Within the dynamic range of 32-bit posits, this figure is unchanged. %
If instead of 16-bits, 32-bits are considered for the proposed system, %
these figures for the worst-case precision jump from 8 to 24. %
Overall, the system achieves a far higher dynamic range %
than both the Posit system and the IEEE-754 system, %
as well as a higher maximum precision in its ideal range %
than both of the counterparts that were considered, %
providing (for example) an additional 4 bits of maximum precision for 32-bit input size %
over the Posit system in its ideal range, and 8 additional bits of precision %
compared to the IEEE-754 system. %
We also present data indicating that the proposed system %
has a range for integer arithmetic on the FPU %
which is larger than that of the Posit system, and also %
that of the IEEE-754 system, for input sizes 32 or larger, %
in spite of a fixed-precision system's natural advantage in this regard. %

In addition to this comparative study, in the main article, we will describe the proposed system in detail, %
develop some of its numerical properties, and make note of some invaluable %
basic results for working in the redundant signed radix~2 %
ternary environment. %

The rest of the contents are as follows. %
In section~\ref{s.basic}, %
basic facts are reviewed, as needed, about the redundant signed radix~2 system. %
In particular, tools are provided for finding units in the last place, %
extrema and ranges for fixed input sizes, and %
the nonadjacent form, both via software and via hardware. %
In section~\ref{s.purereal}, we introduce the proposed system %
in its most basic form, which is the simplest and easiest form to study for the first time. %

In section~\ref{s.extension}, we show that the %
system in the form introduced in section~\ref{s.purereal} %
can easily be modified to achieve %
greater utilization of the available ternary bit fields, %
by providing support for mathematical structures %
needed in some applications, and potential support %
for system artifacts, as we will discuss. %
In effect, we propose in section~\ref{s.extension} a system %
that incorporates not only floating point numbers, %
but also integers, boolean fields, complex floating point numbers, %
as well as real two-vectors and four-vectors, which are important %
in many fields, such as computer graphics and computer vision. %
We show how the system's rich extensibility is provided by a %
re-usable, versatile tool (we refer it to as a point signature) which divides %
the available bit field codes into easily recognized types or classes. %
The incorporation of these elements leads to a system %
that exhibits hardware type safety, %
being closed, out-of-the-box, under a wide range of mathematical and system operations, %
potentially enhancing the robustness, security, and stability of the host system. %

In section~\ref{s.comparison}, %
a comparison of the proposed system with other floating point systems is carried out. %
The Morris system, the IEEE-754 system, and the recent Posit system %
are each reviewed, and then results of an analytically-based comparison of the %
Posit system and the IEEE-754 system with the proposed system %
are reported. %
Source code for the data provided in this section has been made available %
online.\footnote{\url{https://github.com/LuciusSchoenbaum/orencpaper}} %
This section contains convenient formulas giving the %
rounded precision as a function of input real number $x$, %
for each of these three floating point systems. %
Section~\ref{s.conc} is a brief conclusion in which we review 
the advantages and disadvantages of the proposed system. %


\section{The Redundant Signed Radix~2 System}\label{s.basic}

The Booth technique, proposed in 1950 \cite{Booth1}, can speed up operations %
in binary arithmetic by introducing negative-integer-valued digits.\footnote{
This technique can be traced back further to a short paper by Cauchy \cite{CauchySD1}. %
} %
This allows streaks of one's to be eliminated, which reduces the number of %
partial products generated during multiplication. %
A natural extension of the Booth technique is the {\em canonical recoding}, 
also known as the {\em nonadjacent form}, %
defined to be a form in which no %
digit 1 or $\UU$ (negative 1) is adjacent to either a digit 1 or a digit $\UU$. %
We now review this system. 
As needed, theorems in this section are proved in the appendix~\ref{s.proof}. %

\subsection{Basic Statements}\label{ss.statements}
The nonadjacent form of an integer $x$ was noted by Reitweisner to be a unique canonical form  %
with the lowest density of nonzero digits of any %
representation of $x$ in the redundant signed radix~2 system \cite{Reitwiesner1,Koren1}. %
This property has been used in cryptographic applications %
since at least the 1990's, %
see for example \cite{MorainOlivos1,KoblitzCM}. %

Let $x_i$ denote a digit in the digit set $\{1, 0, \UU\}$. %
We define {\em nonadjacent width}, or briefly the {\em width},
of an integer $x$ 
to be the number $n$ %
given by the representation $x = x_n x_{n-1} \dots x_2 x_1 x_0$ %
of $x$ in the nonadjacent form, where $x_n \neq 0$. %
This definition is convenient, for example, in shifting operations. %
The width of $x$ is denoted $\width(x)$. %
By convention, $\width(0) = -\infty$. %

The following statement supplies a handy formula for the nonadjacent width. %

\begin{theorem}\label{t.wd}
The nonadjacent width of an integer $x$ is given by 
\begin{equation}\label{e.rel1}
\width(x) = \floor*{ \log \left(\frac{3|x|}{2}\right) }
\end{equation}
and moreover, 
\begin{equation}\label{e.rel2}
\width(x) = \ceil*{\log \left(\frac{3|x|}{2}\right) } - 1
\end{equation}
since $\log(3x/2)$ is never an integer. %
\end{theorem}

It is sometimes convenient to work with the size %
instead of the width. %
We define the size $\size(x)$ of an integer $x$ to be 
\begin{equation}\label{e.defsize}
\size(x) = 
\begin{cases} 
0 & x = 0, \\ 
\width(x) + 1 & \text{otherwise} 
\end{cases}
\end{equation}
Note that this figure can evidently be regarded as the counting size, measured %
in significant redundant signed digits, of a bit field representation of $x$. %


\begin{theorem}\label{t.jacobsthal}
The number $V_N$ of integers with size $N$, for $N \geq 1$, in their nonadjacent form %
is given by the Jacobsthal sequence (OEIS A001045 \cite{OEIS}). 
That is, %
\begin{equation}\label{e.jacobthal}
V_N = \frac{2^{N+2} - (-1)^{N}}{3}
\end{equation}
When $N$ is even, 
\begin{equation}\label{e.jacobthaleven}
V_N = 4/3(2^N-1) + 1
\end{equation}
\end{theorem}
The following statement is also used often. 

\begin{theorem}\label{t.Xbar}
The largest (smallest, respectively) integer value $\bar{X}_N$ %
($\underbar{X}_N$, respectively) that can be expressed %
in nonadjacent form with a number $N$ of significant digits is
\begin{equation}
\bar{X}_N = \frac{V_N - 1}{2} = \frac{2}{3}\left(2^{N} - \frac{3 + (-1)^N}{4}\right)
\end{equation}
and $\underbar{X}_N = -\bar{X}_N$. 
When $N$ is even, 
\begin{equation}
\bar{X}_N = \frac{2}{3}\left(2^N - 1\right)
\end{equation}
When $N$ is odd, 
\begin{equation}
\bar{X}_N = \frac{2}{3}\left(2^N - 1\right) + \frac{1}{3}
\end{equation}
\end{theorem}

The following re-expression of Theorem~\ref{t.jacobsthal} is sometimes useful. %

\begin{theorem}
For $N \geq 1$, the number $V^1_N$ of positive nonadjacent values %
with exactly $N$ significant digits is
\begin{equation}
V^1_N = V_{N-2}
\end{equation}
They form a closed ball with center 
\begin{equation}
C_N = 2^{N-1}
\end{equation}
and radius 
\begin{equation}
R_N = \bar{X}_{N-2}
\end{equation}
Namely integers $x$ satisfying $C_N - R_N \leq x \leq C_N + R_N$. %
\end{theorem}

The first few of these numbers are given in Table~\ref{t.jacobsthal}. %
Note that although the number of values grows slowly at first, %
the rate of growth quickly becomes exponential. %

\begin{table}[h]
\begin{center}
\begin{tabular}{l|l|l|l|l}
$N$ & $V_N$ & $V_N^1$ & $\bar{X}_N$ & $R_N$ \\ \hline
1 & 3 & 1 & 1 & 0\\
2 & 5 & 1 & 2 & 0\\
3 & 11 & 3 & 5 & 1\\
4 & 21 & 5 & 10 & 2\\
5 & 43 & 11 & 21 & 5\\
6 & 85 & 21 & 42 & 10\\
7 & 171 & 43 & 85 & 21\\
8 & 341 & 85 & 170 & 42\\
9 & 683 & 171 & 341 & 85\\
10 & 1365 & 341 & 682 & 170\\
\end{tabular}
\end{center}
\caption{Count of values, maximum, and radius value for input size $N$.}\label{t.jacobsthal}
\end{table}

%
%

\subsection{Finding the Nonadjacent Form}\label{ss.find}

There are many ways to find the nonadjacent form. %

\begin{example}\label{ex.adhoc}
The first way considered may be performed by hand %
with little difficulty, %
by proceeding from least significant place to most. %
This makes repeated use of the basic relations 
\begin{align}\label{s.nsdbasic}
11 &= 10\UU\\
\UU\UU &= \UU01\\
1\UU &= 01\\
\UU1 &= 0\UU
\end{align}
For example, given %
the following input $x$, we can obtain the nonadjacent form of $x$ %
proceeding from the least significant digit to the most significant digit: %
\begin{align*}
x  &= 1110\UU\UU0\UU0\UU0\UU\UU\UU\UU0010111\\
&= 1110\UU\UU0\UU0\UU0\UU\UU\UU\UU00101\bcancel{1}\bcancel{1}\\
&\phantom{= 1110\UU\UU0\UU0\UU0\UU\UU\UU\UU00101}10\UU\\
&= 1110\UU\UU0\UU0\UU0\UU\UU\UU\UU001100\UU\\
&= 1110\UU\UU0\UU0\UU0\UU\UU\UU\UU00\bcancel{1}\bcancel{1}00\UU\\
&\phantom{=1110\UU\UU0\UU0\UU0\UU\UU\UU\UU0}10\UU\\
&= 1110\UU\UU0\UU0\UU0\UU\UU\UU\UU010\UU00\UU\\
&= 1110\UU\UU0\UU0\UU0\UU\UU\bcancel{\UU}\bcancel{\UU}010\UU00\UU\\
&\phantom{= 1110\UU\UU0\UU0\UU0\UU1}\UU01\\
&= 1110\UU\UU0\UU0\UU\UU0001010\UU00\UU\\
&= 1110\UU\UU0\UU0\bcancel{\UU}\bcancel{\UU}0001010\UU00\UU\\
&\phantom{= 1110\UU\UU0\UU0}\UU01\\
&= 1110\UU\UU0\UU\UU010001010\UU00\UU\\
\end{align*}
We leave the remainder of the steps in this example to the reader. 
\end{example}

\begin{example}\label{ex.software}

Another way of proceeding is to use repeated additions. %
This method is readily implemented in software, see %
Fig.~\ref{f.software}, %
but it %
cannot be considered for hardware implementations, %
because it makes slow, repeated use of addition and branching. %
\end{example}

\begin{figure}[tb]
\begin{betweenlines}
Input: x = x[n-1],x[n-2]...,x[1],x[0]
Output: z = z[n],z[n-1],...,z[1],z[0]
i = 0
while x > 0
	if x is odd
		z[i] = 2 - (x mod 4)
		x = x - z[i]
	else
		z[i] = 0
	x = x >> 1
	i = i + 1
end
\end{betweenlines}
\caption{Software function for nonadjacent form, inappropriate for hardware implementation.}\label{f.software}
\end{figure}

\begin{example}\label{ex.avalues}
A third way of proceeding can be implemented in hardware. %
It shows that the nonadjacent form operation is closely comparable in cost to addition. %
It can be reduced to a single carry chain, and like addition, this %
carry chain is intrinsic to the cost.\footnote{
Algorithms for parallelized addition like \cite{Avizienis1} rely on redundant %
variation in the output. Since the nonadjacent form is unique, they cannot be used for this purpose.}

Given an input $x$, we can calculate the nonadjacent form via a preparing step, %
which generates an associated string of digits with the same length as the input. %
These values may be thought of as a field $a$ of {\em assistant values} $a_{n-1}, a_{n-2}, \dots, a_1, a_0$, or $a$-values. %
The assistant values may be found using Table~\ref{t.avalues}. %
In addition, values referred to as $j$-values are needed, given in Table~\ref{t.jvalues}. %
\begin{table}[tb]
\begin{center}
Row: $x_{n} x_{n-1}$ Column: $a_{n-1}$ Table value: $a_n$\\
\vspace{4pt}
\begin{tabular}{l|lll}
	&1	&0	&\UU \\ \hline
11	&1	&1	&0 \\
10	&1	&0	&\UU \\
1\UU	&0	&\UU	&\UU \\ \hline

01	&1	&0	&0 \\
00	&0	&0	&0 \\
0\UU	&0	&0	&\UU \\ \hline

\UU1	&1	&1	&0 \\
\UU0	&1	&0	&\UU \\
\UU\UU	&0	&\UU	&\UU 
\end{tabular}
\end{center}
\caption{$a$-values}\label{t.avalues}
\end{table}
\begin{table}[htb]
\begin{center}
Row: $x_{n+1} x_{n}$ Column: $a_{n}$ Table value: $j_n$\\
\vspace{4pt}
\begin{tabular}{l|lll}
	&1	&0	&\UU \\ \hline
11	&0	&1	&0 \\
10	&0	&0	&0 \\
1\UU	&0	&1	&0 \\ \hline
01	&0	&0	&0 \\
00	&1	&0	&1 \\
0\UU	&0	&0	&0 \\ \hline
\UU1	&0	&1	&0 \\ 
\UU0	&0	&0	&0 \\
\UU\UU	&0	&1	&0
\end{tabular}
\end{center}
\caption{$j$-values}\label{t.jvalues}
\end{table}

This approach can be reduced to a circuit with only a single carry chain, %
just like one operation of addition, using a repeated array of subunits %
as shown in Fig.~\ref{f.carrychain}. %
Intuitively, the $a$-values produce tags that classify %
the modality of operation in the hand calculation of example~\ref{ex.adhoc}. %
This allows a circuit to perform the entire operation in one pass (a single carry chain). %
A few example calculations are now given. %

\begin{figure}[bt]
\begin{center}
\includegraphics[width=0.4\textwidth]{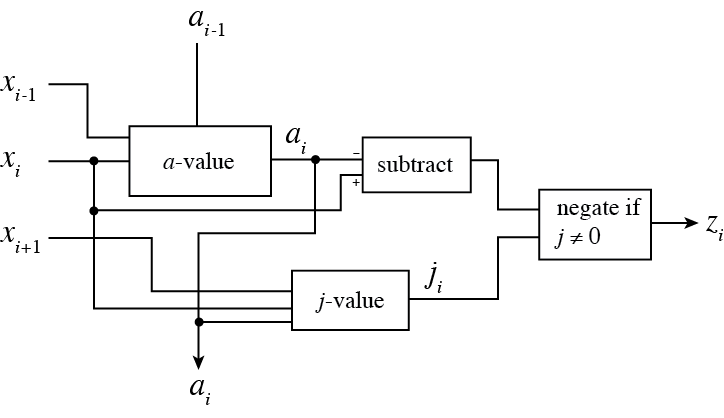}
\end{center}
\caption{Subunit for nonadjacent form (shown: single cell of carry chain array).}\label{f.carrychain}
\end{figure}

If $x = \UU011 \,0111$, its assistant values are 
	$$a = 1111\,1110$$
and the nonadjacent form is 
	$$z = 0\UU00 \,\UU00\UU$$
If  $x = 0\UU11 \,0111$, its assistant values are
	$$a = 0111\,1110$$
the nonadjacent form is 
	$$z = 0000 \,\UU00\UU$$
If $x = 0011 \,0111$, 
the assistant values are
	$$a = 0111\,1110$$
the nonadjacent form is 
	$$z = 0100 \,\UU00\UU$$
For example, if the input $x$ is 
$$x = 00111011000\UU01101110001\UU1\UU\UU\UU0\UU\UU0$$
Then the assistant-values are
$$a = 0111111000011111110000\UU0\UU\UU\UU\UU\UU\UU00$$
and 
$$z = 01000\UU0\UU0000\UU00\UU00\UU0000100001010$$
is the nonadjacent form. 
\end{example}



\section{Real Nonadjacent Forms}\label{s.purereal}

In this section, we introduce the foundational system %
for the proposed floating point system in section~\ref{s.extension}. %
The system we introduce in this section %
encodes nonzero real numbers as floating point numbers, %
and does so in a way that is straightforward and mathematically convenient. %
We refer to this foundational system as the system of {\em real nonadjacent forms.} 
To understand this name, it can be compared to the system of 
{\em complex nonadjacent forms} %
that we introduce in section~\ref{s.extension}. %
The extensions in section~\ref{s.extension} rely on and inherit properties %
from the basic system presented here. %

\subsection{Real Nonadjacent Forms}\label{ss.purereal} 

Consider the nonadjacent forms of two integers $m$ and $n$, 
which are (momentarily) both assumed to be nonzero. %
We can see that, regardless of the length %
of $m$ and $n$'s nonadjacent forms, they could be %
stored in a ternary bit field in a recoverable way %
if we reverse one, say, $n$, and place its most significant ternary bit %
adjacent to the most significant ternary bit of $m$. %
Thus, the nonadjacent form %
allows for a tapered exponent without the need to %
specify the size of the exponent. 
That is, a representation $X$ of a floating point number $x$, 
$$x = m 2^{n}$$
is obtained by reversing the order of the bits of the nonadjacent form of the exponent 
(the integer $n$), 
and concatenating these bits to the front of the nonadjacent form of the significand (the integer $m$), %
understood by convention to be lying in the range $1 \pm \alpha$ or $-1 \pm \alpha$, where %
$\alpha$ is a floating point number determined by the size $N$ (commonly $N = 32$ or $N = 64$) %
of the ternary bit field $X$: %
$$\alpha = \left(\frac{1}{2}\right)^2 + \left(\frac{1}{2}\right)^4 + \dots \left(\frac{1}{2}\right)^k$$
where $k$ is the largest even number $< N$, hence $\alpha \approx 1/3$.\footnote{
Using Theorem~\ref{t.Xbar}, 
$
\alpha = \bar{X}_{N-2} \cdot 2^{-N - 1} 
$.
} %
In other words, the range where precision is at the highest %
is in a range approximately $(0.66, 1.33)$ or $(-1.33, -0.66)$, or $1\pm 1/3, -1 \pm 1/3$. %
In the approach we will adopt shortly below, in this range the precision is fully $N$.\footnote{ %
Note that it is quickly possible from shifting the figure $\alpha$ %
in this way to obtain back-of-the-envelope estimates of precision in various ranges. %
For example, considering only positive values, 
the range where precision is at-worst $N-2$ will be %
$$(4 \times (1 + 1/3), 1/4 \times (1 - 1/3)) = (5.33, 0.16)$$
the range where precision is at-worst $N - 4$ will be %
$$(2^{10_{dec}} \times (1 + 1/3), 2^{-10_{dec}} \times (1 - 1/3)) = (1365.33, 0.00065)$$
A general way to find such ranges is also given, shortly. %
} %

For example, if $m = 1234$, and $n = 5678$, %
and we were to implement this system using decimal digits, %
we would write $X = 87651234$, and we would interpret this %
as the value $x = 1.234 \times 2^{5678}$. %
If we now take two example integers in nonadjacent form, %
say, $m = 10\UU010001$ and $n = 100\UU$, %
we write $X = \UU00110\UU010001$, and interpret this to be the value %
(in signed radix~2) $x = 1.0\UU010001 \times 2^{100\UU}$, %
or in decimal, $(1 - 1/4 + 1/16 + 1/256) \times 2^{8-1} = 0.81640625 \times 2^{7} = 104.5$. %
We can see that from the bit field $X$, we can recover the value, %
regardless of the number of significant digits in $m$ and $n$. %
Thus we have the basis of a tapered floating point system, %
up to the point when either $m$ or $n$ is 0, which we will consider shortly, %
but first, %
we can also see that the unique position where two nonzero values appear, %
is either $1\UU$, $11$, $\UU\UU$, or $\UU1$ depending on %
the signs of the exponent and the significand. %
Thus, these bits are working as efficiently as possible %
to encode useful information, including the sign. %
So there is no need for a hidden bit, or a sign bit, %
and the respective subfields of the field (exponent and significand) %
can be recovered and passed as-is to units equipped to compute in nonadjacent signed radix~2, %
also known as the canonical recoding. %
As we see from the comparisons in section~\ref{s.extension}, %
there is no need for a regime or level-index field %
that would enhance dynamic range. %
In this encoding, round-ties-to-even is implemented by simple truncation. %

Next, we establish conventions for handling values at the extremities 
where either $m$ or $n$ is zero. %
The approach that we adopt is motivated as follows. %
For simplicity of notation, let $|n|$ denote the size $\size(n)$ of $n$. %
We wish to preserve the invariant relation
\begin{equation}\label{e.invariant}
N = |m| + |n|
\end{equation}
for establishing a basis for working with the precision numerically. %
To keep this invariant, we proceed as follows. %
Consider $x = m 2^n$, the same as before, but allowing $m$ or $n$ to be zero. %
We adopt the conventions that:
\begin{enumerate}
\item If the exponent $n$ is zero, the exponent integer $n$ is encoded as an empty field. %
\item The largest size that the exponent may have is $|n| = N-1$, one minus the size of the ternary bit field. %
\end{enumerate}
This pair of rules determines how we interpret a ternary bit field %
that has no pair of adjacent nonzero ternary bits: %
we interpret it as a pure significand, namely $m 2^{0} = m$, %
interpreted as a value lying in the ranges $\pm1 \pm \alpha$, as before. %
To summarize, the value $x$ of the representation %
is obtained by first using the rules above to extract integers $m$ and $n$ 
from the bitfield. Then the value, as a function of $m$ and $n$, is
\begin{equation}\label{e.value}
x(m,n) = m2^{n + |m| - 1}
\end{equation}
To summarize a handful of basic properties, we have:

\begin{theorem}
Each exactly expressible real number $x$ has a unique representation %
in the given encoding. 
If $x$ is exactly expressible, its additive inverse $-x$ is also exactly expressible. %
If $x$ is also a power of two, then its multiplicative inverse $1/x$ is also exactly expressible.\footnote{
This property is also shared by the Posit system, but not by the IEEE-754 system \cite{PositArithmetic}. %
} %
For $x \neq 0$, the exponent $n$ is given as a function of $x$ by the formula
\begin{equation}\label{e.exponent}
n(x) = \width(x)
\end{equation}
Moreover, the unit-in-the-last-place $\ulp(x)$ is given as a function of nonzero real value $x$ %
as 
\begin{equation}\label{e.defulp}
\ulp(x) = 2^{n + |n| - N + 1}
\end{equation}
\end{theorem}

The equation for the $\ulp$ is convenient. %
Using it the Sterbenz lemma can be verified. %

From the invariant relation Eq.~\ref{e.invariant}, we have: %

\begin{theorem}\label{t.precision}
For a real number $x \neq 0$, and choice of input size $N > 0$, let 
\begin{equation}\label{e.precision}
p = N - \size(\width(x))
\end{equation}
Then the precision in which $x$ is represented %
in the proposed system is exactly $p$, %
and $x$ is within the expressible range of the system (i.e., the dynamic range) if and only if %
$p > 0$. %
\end{theorem}
From Theorem~\ref{t.precision} we can consider precision %
as an elementary function of the value $x$. %
Since it may aid the reader's sense of balance in the freshly proposed system, %
a toy example system $N = 4$ is explicitly enumerated in Fig.~\ref{f.N4}. %
Encodings appear on the left, and corresponding values appear on the right. %
Horizontal bars indicate where an ulp shift occurs. %
\begin{figure}
\begin{center}
\begin{tabular}{cc}
\begin{tabular}{c}
\begin{tabular}{cc}
\begin{tabular}{l}
$\UU0\UU1$\\
$00\UU1$\\
$10\UU1$\\ \hline
$0\UU10$ \\ \hline
$\UU10\UU$ \\
$\UU100$ \\
$\UU101$ \\ \hline
$10\UU0$ \\
$100\UU$
\end{tabular} &
\begin{tabular}{l}
$1011$ \\
$0011$ \\
$\UU011$ \\ \hline
$0110$ \\ \hline
$1101$ \\
$1100$ \\
$110\UU$ \\ \hline
$1010$ \\
$1001$
\end{tabular}
\end{tabular}\\
\begin{tabular}{l}
$1000$\\
$\UU000$
\end{tabular}\\
\begin{tabular}{cc}
\begin{tabular}{l}
$\UU001$ \\
$\UU010$ \\ \hline
$\UU\UU0\UU$ \\
$\UU\UU00$ \\
$\UU\UU01$ \\ \hline
$0\UU\UU0$ \\ \hline
$10\UU\UU$ \\
$00\UU\UU$ \\
$\UU0\UU\UU$
\end{tabular} &
\begin{tabular}{l}
$\UU00\UU$ \\
$\UU0\UU0$ \\ \hline
$1\UU01$ \\
$1\UU00$ \\
$1\UU0\UU$ \\ \hline
$01\UU0$ \\ \hline
$\UU01\UU$ \\
$001\UU$ \\ 
$101\UU$
\end{tabular}
\end{tabular}
\end{tabular}
&
\begin{tabular}{c}
\begin{tabular}{cc}
\begin{tabular}{r}
$1/32$ \\ 
$1/16$ \\
$1/8$ \\ \hline
$1/4$ \\ \hline
$3/8$ \\
$1/2$ \\
$5/8$ \\ \hline
$3/4$ \\
$7/8$
\end{tabular} &
\begin{tabular}{r}
$32$ \\
$16$ \\
$8$ \\ \hline
$4$ \\ \hline
$5/2$ \\
$2$ \\
$3/2$ \\ \hline
$5/4$ \\
$9/8$
\end{tabular}
\end{tabular}\\
\begin{tabular}{r}
$1$\\
$-1$
\end{tabular}\\
\begin{tabular}{cc}
\begin{tabular}{r}
$-7/8$ \\
$-3/4$ \\ \hline
$-5/8$ \\
$-1/2$ \\
$-3/8$ \\ \hline
$-1/4$ \\ \hline
$-1/8$ \\
$-1/16$ \\
$-1/32$
\end{tabular} &
\begin{tabular}{r}
$-9/8$ \\
$-5/4$ \\ \hline
$-3/2$ \\
$-2$ \\
$-5/2$ \\ \hline
$-4$ \\ \hline
$-8$ \\
$-16$ \\
$-32$
\end{tabular}
\end{tabular}
\end{tabular}
\end{tabular}
\end{center}
\caption{Proposed floating point system for input size $N = 4$.}\label{f.N4}
\end{figure}
The reader may also wish to verify %
that the system for $N = 2$ has six exactly expressible values (excluding zero): %
one, two, $\frac{1}{2}$, and their negatives. %

\subsection{Rounding}\label{ss.round}

It is now possible to discuss rounding in the present system. %
For this, we use the precision $p = p(x)$ introduced in the last section. %
We also let $\Omega_N$ denote the largest exactly expressible value in the %
system, namely 
$$\Omega_N = 2^{\bar{X}_{N-1}}$$ %
Still fixing $N > 0$ and $x$ as before, let $R(x)$ be the rounded %
expression of $x$. %
We can define $R(x)$ unambiguously for all $x \neq 0$ as
\begin{equation}\label{Rdef}
R(x) = \begin{cases}
\hat{x} \, 2^{\width(x)} & p(x) > 0 \\
\sgn(x) \,\Omega_N & p(x) \leq 0, \width(x) > 0 \\
\sgn(x) \,\Omega_N^{-1} & p(x) \leq 0, \width(x) < 0
\end{cases}
\end{equation}
where the expression $\hat{x}$ denotes the %
value of $x$ shifted so it lies in one of the ranges $\pm 1 \pm \alpha$, %
and then truncated at the length $p(x)$ in the canonical recoding. %
This routine for finding $R(x)$ implements the round-ties-to-even %
rounding policy, since %
once the expansion in nonadjacent form of $x$ is found, %
rounding ties to even is performed via simple truncation. %

This rounding policy is identical to that of the recently proposed Posit system. %
In particular, finite numbers are never rounded up or down to infinity, %
and nonzero numbers are never rounded to zero. %
However, the rounding method itself is slightly different for each system. %
In the Posit system, there is an exceptional %
case which arises when the rounded bit is not a fraction bit \cite{PositArithmetic}. %

The IEEE-754 standard uses a different rounding policy, %
in particular it rounds values higher than its maximum value $\Omega$ %
to infinity $+\infty$, as an overflow value. %
This is done in order to provide for a guaranteed bound %
on the relative error on a result whose value is $\Omega$. %
In the present system, the order of magnitude of $\Omega$ in %
32-bit single precision, for example, is %
over a million times larger than that of $\Omega$ %
for the IEEE-754 system. %
So it can be considered effectively infinite for practical purposes. %
In other words, a calculation whose result is $\Omega$ is already %
an indication that something is wrong, %
and it suffices to have error bounds for all other finite values, including $\Omega/2$, %
the penultimate representable value. %




\section{Extensions of the Basic System}\label{s.extension}

In section~\ref{s.purereal} we introduced essentially a tapered floating point %
system, and we enumerated some of its basic properties. %
In this section we extend the system introduced in section~\ref{s.purereal}, %
which has underutilized the possible encodings available to its underlying ternary bitfield, %
to make better use of the available encoding space. %
We introduce a concept called a {\em point}, and use this to extend %
the system to incorporate complex values, and other types. %
The underlying %
point concept is analogous to a decimal point %
just like that of schoolhouse mathematics. %
Thus, the system is simple, and convenient to implement and maintain. %

\subsection{Points and Real Nonadjacent Forms}\label{ss.point}

In this section, we make the only (very small) breaking %
change in the family of encodings that we discuss here. %
For the convenience of the reader, %
we distinguish the breaking change within the nomenclature. %
From this section on, the {\em real nonadjacent forms} %
will refer to the encoding we present here, %
not that of section~\ref{s.purereal}, %
which is referred to as the {\em pure real nonadjacent forms}, %
or something to that effect. %
When studying the properties of the proposed system as a pure floating point system, %
it is sometimes convenient to just work with the {\em pure real} nonadjacent forms %
of section~\ref{s.purereal}, 
while an implementation using the {\em real} nonadjacent forms, %
introduced now, will benefit from enhanced extendability, as we see shortly. %

The reason for the breaking change is that 
the proposed extensions will all be based on one new notion, that of a {\em point}. %
Given a ternary field $x$, a {\em point} is defined to be %
a contiguous subfield, of length 2 or greater, in which no zeros %
appear, and %
which is bounded on both the left and right either by a zero, %
or by the left or right boundary of the bit field. %
For example, the fields $0110$ and $1\UU1\UU$ each contain one point, %
and the field $1\UU01\UU\UU110101\UU\UU$ contains three points. %
We should already have some intuition for this notion, %
from our previous work in section~\ref{s.purereal}. %

Every ternary-bit field has a well-defined positive or zero %
number of points, according to this definition. %
It is clear that any integer expressed in nonadjacent form (section~\ref{s.basic}) %
has zero points. %
We can also see that most (but not all) pure real forms (section~\ref{s.purereal}) have exactly one point. %
However, the case when the exponent $n$ is zero is another case where %
the number of points drops again to zero. %
So in that case, the field becomes indistinguishable %
from an integer in nonadjacent form. %

We would like to correct this: %
we would like for all the pure real %
forms to have exactly one point. %
The purpose of this is not to make room for the integers in nonadjacent form%
---although that is a side effect. %
Rather, the purpose is to be able to install a valuable convention, namely, %
if there are exactly two points found in the expression, %
then by convention, the field is cut exactly into two equal parts, and those parts %
are regarded separately as fields having one point each. %
This convention may be implemented recursively\footnote{
As a part of this recursion, %
it will always be assumed that the points divide evenly %
among the subfields, once the field is divided. %
Otherwise, the field has no assigned meaning. %
We reserve the right to make this assumption from now on, %
without necessarily commenting about it explicitly. %
} %
as we explore %
further in section~\ref{ss.pointsignature}. %

We can do this if we add a small map to the %
specification of the pure real system. %
Namely, we remap the {\em second} and {\em third} most significant ternary bit of the significand %
to a different bit-level form. %
The specification is as follows. %
Let $X$ be a bitfield representing a floating point number with exponent zero in the pure real form %
of section~\ref{s.purereal}. %
Then this class of $X$ is precisely that of the pure real forms having zero points. %
For convenience, suppose (without loss of generality) that $X$ represents a positive real number $x > 0$. %
Write
$$X = 1x_1 x_2...$$
where $x_1 x_2$ denotes the ternary bits in the second and third most significant places, 
so that this 1 that appears is the most significant digit. %
We can see that $x_1 x_2$ has the form $0 t_0$, where $t_0$ is an arbitrary ternary bit, %
either $1, 0$, or $\UU$. %
We can therefore map the field $x_1 x_2$ one-to-one to a corresponding field $y_1 y_2$, as in: %
$$X = 1y_1 y_2...$$
This creates a single point in the bitfield $X$, as desired. %
We call this the {\em point correction mapping}. See Table~\ref{t.pointcorrection}. 

\begin{table}[htb]
\begin{center}
\begin{tabular}{c|c}
$x_1 x_2$ & $y_1 y_2$ \\ \hline
$\raisebox{2pt}{\strut}
01$ & $\UU1$ \\
$00$ & $\UU\UU$ \\
$0\UU$ & $1\UU$
\end{tabular}
\end{center}
\caption{Point correction mapping}\label{t.pointcorrection}
\end{table}

There is an unused/reserved case, when $X = 1110...$ or $X = \UU110...$ %
We call these remaining cases {\em point-escaped} values. %
We define two such values, plus infinity ($+\infty$) and minus infinity ($-\infty$). %
Each of these are identified by the local bits alone, using the properties: 
\begin{itemize}
	\item positive infinity ($+\infty$), $X = 111...$: the field has one point, and starts on the left with the bit pattern $111$. 
	\item negative infinity ($-\infty$), $X = \UU11...$: the field has one point, and starts on the left with the bit pattern  $\UU11$. 
\end{itemize}
A glaring special case, that we have put off until now, is the value zero. %
It is possible to use a point-escaped field to define the value zero. %
However, in order to avoid introducing another exceptional value to the system, %
we prefer that the encoding for the value zero is the %
field of zeros, the same encoding as in the nonadjacent integer system. %
This makes sense, because a representation of zero in a floating point system %
is a symbolic entity (not a numerical one), %
and the system of integers is a field of symbolic entities which already contains 
a symbolic zero. %

Thus, the system shares in common with the IEEE-754 system %
and the Posit system the property that zero (or more precisely $+0$, in the IEEE-754 case) %
is encoded with the field of zeros. %

The point-escaped values $\pm\infty$ are defined %
for symbolic purposes (for example, division of a nonzero floating point number by symbolic zero) %
as symblic infinities can arise %
in useful (symbolic) results in many contexts. %
It is suggested, then, to regard $\pm\Omega$ in the nonadjacent system %
as {\em numerical infinite} values, while $\pm\infty$ in the nonadjacent system %
can be regarded as {\em symbolic infinite} values. %

This completely describes the system of {\em real nonadjacent forms}, %
which are also understood to contain the representations of integers %
in their nonadjacent forms. %

\subsection{Boolean Forms}\label{ss.boolean}

Before we go further, we note that the system can be regarded as %
containing a further extension, to {\em boolean forms}. %
Namely, we can map the value 1 to the boolean value {\bf true}. %
and $\UU$ to the boolean value {\bf false}. %
Then ternary bit fields in which no zero appears %
can be interpreted as boolean fields, %
which are distinguishable from all of the real nonadjacent forms 
of section~\ref{ss.point}. %

In particular, the field of all ones may be read as {\bf true}, %
boolean true, and the field of all $\UU$ may be read as {\bf false}, %
boolean false. %

\subsection{Complex Nonadjacent Forms}\label{ss.complex} 

Just above, we introduced the notion of a {\em point}. %
We noted that every ternary bit field has a positive or zero number of points. %
We modified the system of section~\ref{s.purereal} slightly, so that
each encoded tapered floating point representation, including exceptional values, has at most one %
point, %
out of the possible range of points $11, 1\UU, \UU1, \UU\UU$ (points with length 2), %
$111, 11\UU, 1\UU1, 1\UU\UU, \UU11, \UU1\UU, \UU\UU1, \UU\UU\UU$ (points with length 3), %
and also points of length 4, however points of length 3 or 4 are only possible if they are %
on the leftmost side of the bit field. %
As we noted, %
this allows us to represent integers, boolean fields, and real numbers %
in the same system. %
However, we have not come close to exhausting the possible encodings. %
Now, we wish to go further and provide an encoding for complex numbers. %

To encode the complex numbers, %
we utilize the case when the bit field contains %
two points, in such a way that %
when the field is divided down the middle into $N/2$ %
equal subfields, each subfield has exactly one point. %
We associate to the left subfield a floating point number %
whose value is $a$, and the right subfield %
a floating point number whose value is $b$, %
and we interpret the full ternary bit field %
as representing the quantity $a + bi$, %
where $i$ is the complex unit $i = \sqrt{-1}$. %

This describes the extension, %
up to what is done 
when either $a$ or $b$ is $+\infty$ or $-\infty$, %
or when either $a$ or $b$ is zero. %
For the infinite cases, we handle these by the rule:
\begin{enumerate}
	\item If a field contains exactly two points and begins with the pattern $111...$, %
	then it represents the value {\em complex infinity}, denoted $\infty$. %
\end{enumerate}
Next to consider is the zero cases. %
First, notice that if we try, in the na\"ive way, %
to represent either $0 + bi$ or $a + 0i$ in the encoding %
scheme proposed shortly above (in which we %
divide the field into two equal subfields), %
and use the encoding of zero established in the preceding section, %
we obtain a bit field with only one point, not two points, %
meaning that it is indistinguishable from a real nonadjacent form. %
We also make note of the possibility that a real nonadjacent form can be %
represented in two ways, either as real $a$, or complex $a + 0i$, %
each with a different precision, %
which threatens to add a clutter of case-switching to the resulting system. %
We can therefore consider the possibility of %
making the complex numbers a true extension of the preceding encoding, %
and thereby avoid such clutter to unburden potential hardware implementations. %

To do this, we now wish to introduce another type of point. %
The points defined in section~\ref{ss.point} %
are now {\em Type I} points, and %
we now wish to define {\em Type II} points. %
To do this, we introduce a generalization of the point correction mapping. %
Instead of a single point correction mapping, %
we now have a family of five them, labeled %
PCM1, PCM2, PCM3, PCM4, and PCM5. %
The PCM1 mapping is the original point correction mapping %
in Table~\ref{t.pointcorrection}, associated with Type I points. %

We do not fully specify the PCM2, PCM3, PCM4, or PCM5 mappings at the bit level %
here, as it may be technology-dependent and is not really relevant here. %
However, 
we can provide a sketch of them. %
The PCM2 mapping is as follows: %
given a point of length 2 somewhere in the middle of the bit field, as in
$$X = ...0 p_1 p_2 0...$$
we modify the point by changing the right zero to a nonzero digit $y_1$:
$$X = ...0 p_1 p_2 y_1...$$
This gives a point of length either 3 or 4. %
The PCM3 mapping changes a point of length 2 %
positioned at the left of the bit field, as in:
$$X = p_1 p_2 0 x_1 x_2...$$
to a point of length either 5 or 6 by mapping the pattern $0 x_1 x_2$ to a %
pattern $y_1 y_2 y_3$ with no zeros:
$$X = p_1 p_2 y_1 y_2 y_3...$$
There are still two more, PCM4, and PCM5. %
PCM4 is similar to PCM3, except it takes a point of length 2 positioned at the right of the bit field, as in:
$$X = ...x_1 x_2 0 p_1 p_2$$
and maps it to a point of length either 5 or 6, as in:
$$X = ...y_1 y_2 y_3 p_1 p_2$$
Finally, PCM5 takes a bit field with no point (encoding a real %
number with the maximum precision), as in 
$$X = p_1 0 x_1 x_2 x_3 x_4 x_5...$$
and replaces the pattern $0 x_1 x_2 x_3 x_4 x_5$ with nonzero digits 
$y_1 y_2 y_3 y_4 y_5 y_6$, as in:
$$X = p_1 y_1 y_2 y_3 y_4 y_5 y_6...$$
giving a point of length either 7 or 8. %

We can now see that we are able to recognize 
what PCM mapping to apply, based on %
\begin{enumerate}
	\item whether a point is adjacent to the left of the bitfield (case L), %
	adjacent to the right of the bitfield (case R), %
	or otherwise (case M),
	\item the length of the point, either 2, 3, 4, 5, 6, 7, or 8. 
\end{enumerate}%
We briefly call this information the {\em point class}. %
The point class provides us with the ability to recover the %
original point, and to derive the point type. %
This information is summarized in Fig.~\ref{f.type12points}. 
In the figure, PCM0 denotes no modification to the point. %

\begin{figure}[thb]
\begin{center}
\begin{tabular}{c|l|l|l}
visual aid	& point class 	& point type 	& PCM \\ \hline
\begin{tabular}{c}
$...0pp0...$ \\
$...0pp$\\
$pp0...$
\end{tabular}	& 
\begin{tabular}{l}M2 \\ R2 \\ L2 \end{tabular}	
						& I			& PCM0 \\ \hline
$p0...$		& L3, L4		& I			& PCM1 \\ \hline
$...0pp0...$		& 
\begin{tabular}{l}
M3, M4, \\ R3, R4 \end{tabular} & II			& PCM2 \\ \hline
$pp0...$		& L5, L6			& II			& PCM3 \\ \hline
$...0pp$		& R5, R6			& II			& PCM4 \\ \hline
$p0...$		& L7, L8			& II			& PCM5
\end{tabular}
\end{center}
\caption{Definition of point types: type I, type II.}\label{f.type12points}
\end{figure}

Now we can return to the complex number encoding. %
Equipped with two types of point, we proceed by cases given a complex number $z$:
\begin{enumerate}
\item
If $z$ is pure real, it is encoded as a real number using a type I point. %
\item
If $z$ is pure imaginary, it is encoded as a real number using a type II point. %
(Note the agreement when $z$ is both pure real, and pure imaginary.) 
\item 
Otherwise, $z = a+bi$ for two nonzero $a$ and $b$. %
These real numbers $a$ and $b$ are encoded as real nonadjacent forms %
in the size $N/2$, and encoded together in concatenated form, using a type I point %
for $a$ and a type II point for $b$, in a bit field of size $N$.\footnote{
It is a choice to use a type II point for $b$. It may be %
amenable to further generalizations to proceed this way. %
} %
\end{enumerate}

This provides for an extended number system that we call the {\em complex nonadjacent forms}. %
This system contains the real nonadjacent forms as a precise, one-to-one subset. %
This system preserves the property of the real nonadjacent forms, %
that there is only one way to represent zero (as the field of zeros), %
regardless of whether it is complex, real, or an integer. %
It contains three exceptional values: $+\infty, -\infty, $ and the complex point at infinity (projective infinity) $\infty$. %

Note that, to preserve the type safety established for integers, real numbers, and booleans %
in section~\ref{ss.point} and section~\ref{ss.boolean}, %
we can only use type~II points in fields of length $N > 8$. %
So the complex nonadjacent forms would require a minimum standard input size %
$N = 32$. 

%

\subsection{Point Types and Point Signature}\label{ss.pointsignature}

In fact, the utilization of the system is still far below the maximum. %
To make further use of the unused encoding space, %
we can make further use of point-escaped values. %
We can also make further use of point types. %
It is easy to do this, in fact (at the level of specification), %
if we consolidate this additional structure with the notion of a point signature. %

If we scan a ternary bit-field from left to right, %
we can make a list of the points that we observe. %
By convention, we write this list in the form $\{pt_1, pt_2, \dots, pt_k\}$, %
where each variable $pt_i$ denotes an integer that %
expresses what type of point is observed (either type I or type II), %
and $k$ is the number of points observed, or $1$ %
if no points are observed. 
Note that we also say that a {\em full point} %
is a point that has the full length of the bit field. %
The following scheme includes all exceptional cases: %
\begin{enumerate}
\item $pt_i = 1$: a type I point is detected. 
\item $pt_i = 2$: a type II point is detected. 
\item $pt_i = \infty$: a full point is detected.
\item $pt_1 = 0$, no point is detected, and the field is not all zeros. %
\item $pt_1 = -1$, the field is all zeros. %
\end{enumerate}
The resulting {\em point signature} %
is a succinct way to %
express extensions to the basic system of real and complex nonadjacent forms. %

With this, we can succinctly recapitulate all the material of %
this section:
\begin{enumerate}
\item 
If the field has a point signature $\{\infty\}$, %
we consider it to be a {\em boolean}. 
\item
If the field has a point signature $\{0\}$ or $\{-1\}$, %
we consider it to be an {\em integer}. %
\item
If the field has a point signature $\{1\}$ or $\{-1\}$, 
we consider it to be an {\em extended real number}: either a finite real number, $+\infty$, or $-\infty$. 
\item
If the field has a point signature $\{1\}$, $\{2\}$, $\{1,2\}$, or $\{-1\}$, %
the field is considered to be a {\em complex number}, or else complex infinity $\infty$.
\end{enumerate}
Other than the exceptional field of only zeros, %
which is regarded as simultaneously a real, an integer, and a complex number, %
and the signature $\{1\}$ which is either %
real or complex depending only on the context, %
the classification is a division %
of the possibilities into distinct classes. %

\subsection{Vector Nonadjacent Forms}\label{ss.vec}

We can use point signatures to define further extensions of the system for vector forms. %
If the field has a point signature $\{1,1\}$, %
the field may be considered to be a two-vector, 
that is, a pair of reals. %
We can also reserve the case of point signature $\{1, 1, 1, 1\}$ %
for four-vectors.\footnote{
As in the previously encountered cases, %
there is no assigned meaning unless %
the bitfield can be divided in half in such a way that 
the point signatures divide evenly. %
For example, dividing a bit field with point signature $\{1, 1, 1, 1\}$ %
in half must give two bit fields with point signature $\{1, 1\}$ %
and these in turn must both divide into two equally-sized bit fields with point signature $\{1\}$. %
} %
A further unused case is $\{1, 2, 1, 2\}$, which %
might be considered as a foundation for complex two-vectors. %
A further possibility is to encode higher number fields, %
such as quaternions and octonions, %
by making use of unused point signatures. %
One possibility is to use the signature $\{1, 1, 1, 2\}$ %
as a foundation for a quaternion encoding. %

In raising the possibility of %
these extensions, no mention has been made about exceptional cases, including zero cases. %
We do not delve further into possible ways of handling these exceptions here, %
as the proposed extensions are not well tested at the present time, 
and are only offered on a preliminary basis, along with a call for comments. %

%

\begin{figure}[htb]
\begin{center}
\includegraphics[width=0.5\textwidth]{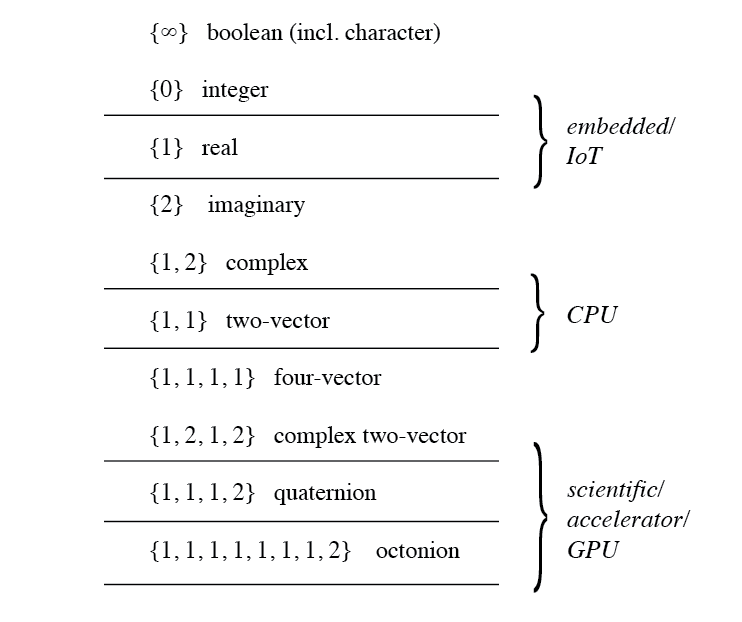}
\end{center}
\caption{Point signatures, with vertical lines indicating selected implementation finish lines.}\label{f.pointsignature}
\end{figure}

We give a summary of the point signatures %
in the proposed system in Fig.~\ref{f.pointsignature}, 
where the zero case is omitted for clarity. %
In the figure, the vertical lines indicate divisions %
where implementations might stop providing hardware support, %
along with a suggested hardware class in each case, among %
embedded/IoT devices, CPU class devices, and scientific/accelerator devices, including GPUs. %

\subsection{Instruction Types}\label{s.security}

The proposed system exhibits a form of hardware type safety, %
namely, various mathematical objects (booleans, integers, real and complex numbers, etc.) %
are represented in a common encoding. %
We now make brief mention of the possible implications of this property. %

Considering that character encodings can all be assimilated into the boolean category, %
the potential arises for a firm, system-wide line between code and data, %
which could be drawn via the further use of the point signature notion (and possibly, %
further use of point correction mappings to encompass further point types). %
Using only this technique, it would be possible %
to encode machine instructions in a such a way that their machine and memory representations %
are always immediately distinguishable from data, 
which would eliminate a range of security attacks. %
Support for this feature would have to encompass controls on the ISA %
(instruction set architecture, including the machine instruction set), %
whereby operations would be required to be closed operations on the data types, %
simply so that it is impossible to transform data into machine instructions %
that could be run to malicious ends. %
For example, it would be harmless to allow boolean operations %
on boolean types, as long as via such operations, %
it is not possible to transform boolean types into other types 
(real numbers, integers, machine instruction types). %
This implies that, if such a feature is targeted, %
the full family of ternary operations that would be present at the hardware/circuit level %
would not be exposed at the level of the ISA/architecture. %
This is perhaps reason for some relief, as the number of ternary boolean %
operations numbers well into the thousands.\footnote{
There are $\nondegenterngates$ nondegenerate ternary-valued functions of two ternary inputs. %
Of these, there are $\nondegenterngatescmu$ commutative functions, and $\nondegenterngatesnoncmu$ noncommutative functions. 
} %


\section{Comparisons With Morris, IEEE-754, and Posit Systems}\label{s.comparison}

In this section we compare the proposed system to %
other systems that have been proposed. %
We consider three systems: the original tapered system due to Morris \cite{Morris1}, %
the IEEE-754-1985/2008 system \cite{IEEE754_1985, IEEE754_2008}, 
and the recently proposed Posit system \cite{GustYo1,PositStandardDocs3-2}. %

Throughout this section $x$ denotes a nonzero real number. %
For each system, we will consider the precision %
of the corresponding representation %
as a function of the exponent in base 2 of a represented value $x$, %
or the {\em precision by order of magnitude} (PBOM). %
Note that PBOM factor considers order of magnitudes in base~2, %
not in base~10, as is common in many fields. %
We denote this factor briefly by $B$, or $B_N$. %
We denote the base-2 order of magnitude %
of a represented value $x > 0$ by $n_x$:
$$
n_x = \floor*{\log_2 x}
$$
So we can write $B_N$ as $B_N(n_x)$ as it is %
a function of the order of magnitude $n_x$ of represented value $x$. %

\subsection{Overview}\label{ss.overview}

In this section, we summarize in condensed form some of the most %
salient features of each floating point encoding system. %

\subsubsection{Fractional part encoding scheme} 
The Morris system, the IEEE-754 system, and the Posit system %
all utilize a hidden bit and a sign bit for the significand/fraction. %
The proposed nonadjacent system instead stores all of the value's sign %
information in the point. %
\subsubsection{Exponent encoding scheme} 
For the exponent, the Morris system uses sign-magnitude encoding. %
The IEEE-754 system uses a concept called bias that is %
related to subnormals: %
the bias value is where the subnormal exponent range will start. 
The Posit system uses a mixed-base encoding, one subfield using signed \mbox{base-1}, %
and the remainder using unsigned \mbox{base-2}. %
\subsubsection{Tapering}
All of the systems use some form of tapering. %
IEEE-754 uses a fixed-size exponent in the normal range, 
while in the subnormal range (around zero), the precision tapers off %
and contributes instead to the exponent (in \mbox{base-1} fashion), %
to allow for gradual underflow. %
The Morris and Posit systems use a tapered exponent (variable size exponent), %
and so does the proposed nonadjacent system, %
but each system utilizes its own %
distinct mechanism to achieve the characteristic tapering effect. %
\subsubsection{Exponent} The Morris system uses a fixed-size subfield $G$ to represent the tapered exponent length. %
The choice of the size $g$ of $G$ is arbitrary and must be chosen by the designer. %
Guidance for choosing $g$ was not implemented by Morris. %
The IEEE-754 system uses a fixed-size exponent field, %
which has well-known advantages for hardware. %
The choice of the size of the exponent field is arbitrary and could be chosen by the designer, %
but the standard has well-established guidelines on the choice for standard input sizes. %
The Posit system uses a fixed-size subfield for part of the exponent, %
however, it uses a built-in convention that allows the subfield to be arbitrarily truncated on the right. %
The choice of the size $es$ of this subfield is arbitrary and must be chosen by the designer, %
but the Posit system proposal \cite{PositStandardDocs3-2} establishes recommended/standard values of $es$ for %
the standard input sizes. %
The Posit system uses a \mbox{base-1} field to encode the %
major contribution in the exponent (the regime). %
\subsubsection{Rounding} 
The Morris system gives no recommendations/suggestions for rounding policy. %
The IEEE-754 standard provides four rounding conventions: %
round-up, round-down, round-to-zero, and round-ties-to-even, 
with round-ties-to-even as the default policy. %
For values greater than the dynamic range, %
it rounds to $+\infty$. %
Posits use round-ties-to-even only, %
and never rounds non-representable values to zero or infinity. %
\subsubsection{Exceptions} 
The Morris system specifies no exceptional values. %
The IEEE-754 system has several well-known exceptional values, %
including NaN (with payload), $\pm0$, and $\pm\infty$. %
The Posit system defines two exceptional values, 0 and a value NaR, which %
englobes all infinite and undefined quantities. %

Next, we will provide a description of each of the systems, %
and supply the PBOM factors for the Posit system %
and the IEEE system. %

\begin{figure*}[thb]
\begin{center}
\includegraphics[width=0.98\textwidth]{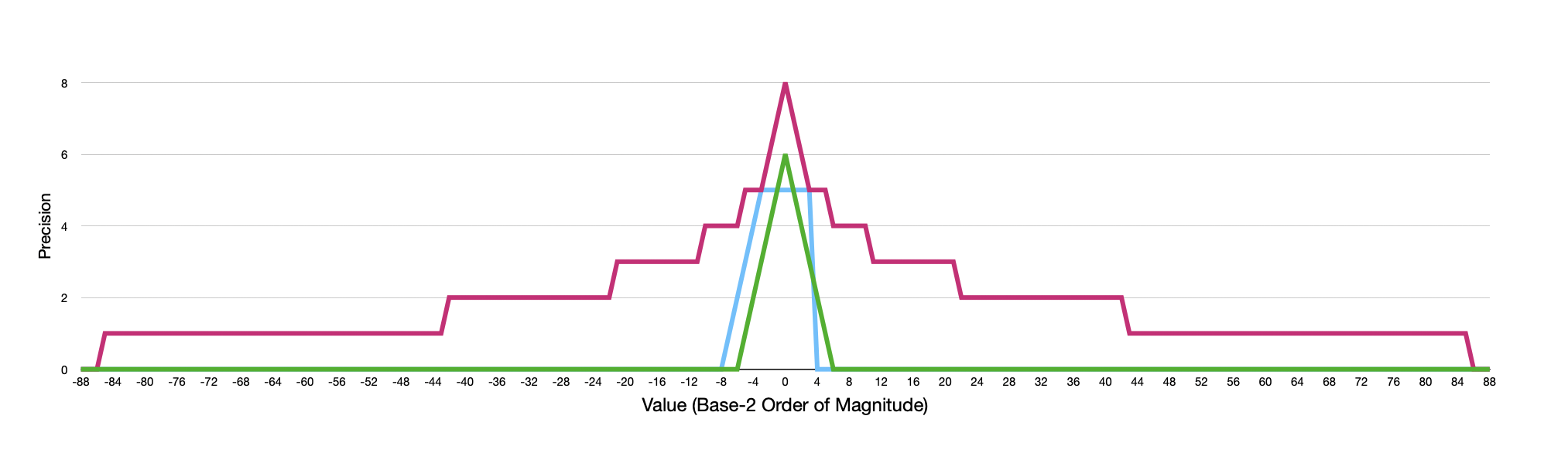}
\end{center}
\caption{Precision by order of magnitude (PBOM) for the three systems for ultra-low precision input size $N=8$ showing IEEE in blue, POSIT in green, NONADJ in purple.}\label{f.BcurvesN8}
\end{figure*}

\begin{figure*}[thb]
\begin{center}
\includegraphics[width=0.98\textwidth]{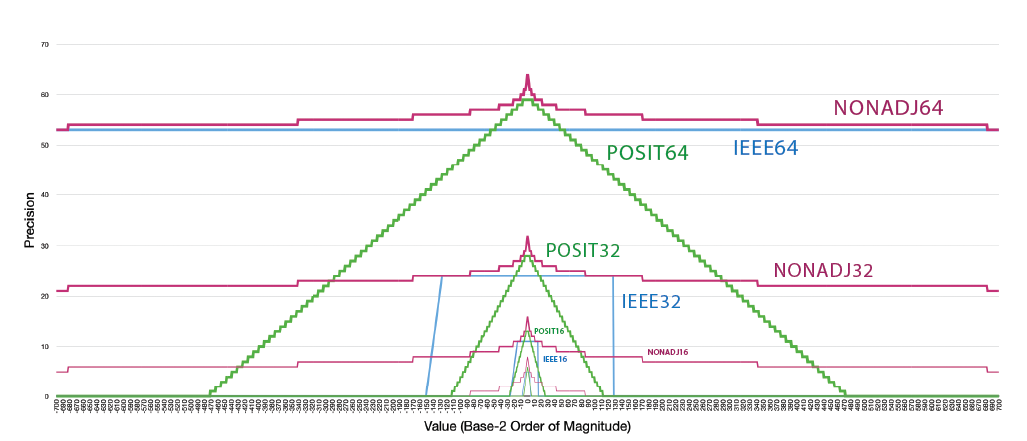}
\end{center}
\caption{Precision by order of magnitude (PBOM) for the three systems IEEE, POSIT, NONADJ, %
input sizes $N=8, N=16, N=32, N=64$, in range from $-700$ to $700$.}\label{f.Bcurves}
\end{figure*}


\subsection{Morris Tapered Numbers}\label{ss.morris}

The proposed number system of Morris \cite{Morris1} %
is the earliest proposal of a tapered number system %
for use in computing machines. %
Its characteristic feature is a fixed length subfield of size $1 \leq g_N \leq \bar{g}_N$ %
where $\bar{g}_N$ is a maximum value $\leq N$, %
which presents a figure $G$ as an unsigned binary integer. %
Then, it provides for two variable-length fields that present %
two unsigned integers $m$ and $n$, with the size of the field for $n$ %
being always exactly $G+1$. %
Finally, it has two reserved bits for the signs $s_n$ and $s_m$ of $m$ and $n$, %
respectively, so that $n$ and $m$ are essentially quantities encoded in %
sign-magnitude, as in the illustration: 
$$
\includegraphics[width=0.4\textwidth]{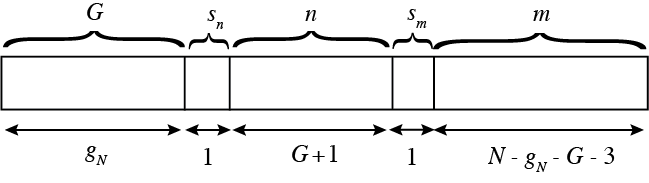}
$$
From this the maximum value $\bar{g}_N$ of $g_N$ is the largest value to satisfy
\begin{equation}
g_N + 2^{g_N} +2 \leq N
\end{equation}
No bias is considered, and the encoded value is defined as follows: 
\begin{equation}
x = (-1)^{s_m} \cdot (1 + m \cdot 2^{-\log_2 (m) -1 } ) \cdot 2^{(-1)^{s_n} \cdot n}
\end{equation}
where the convention is $0 \cdot 2^{\infty} = 0$. %
The dynamic range is then %
\begin{equation}
2^{\pm 2^{g_N}}
\end{equation}
Morris suggests $g_{36} = 3$. %
Morris considers $N=36$, having in mind the IBM 7090 mainframe computer %
which uses a fixed 7 bits for the exponent and 27 bits for the fraction, plus two sign bits. %
In the Morris system many values are encoded redundantly in multiple different ways, %
but as noted by Morris, this can be avoided if the size $G+1$ of the field for $n = n_x$, %
the order of magnitude of represented value $x$, %
is taken to be the size $1 + \log_2 n_x$ of $n_x$ when $n_x \neq 0$. %
In that case, the precision is %
\begin{equation}
p = N - g_N - 2 - \log_2 n_x
\end{equation}
but the case of exponent zero ($n_x = 0$) is exceptional. %
In this system, the maximum dynamic range can be considerably larger %
than for a fixed-point system, %
at the cost of precision across the entire dynamic range. %
We do not study the precision by order of magnitude (PBOM) %
for the Morris system here, %
but if we did, the curve would have a shape like that of the nonadjacent system. %
This is because the Morris system, like the nonadjacent system, %
encodes the exponent in \mbox{base-2} fashion. %

\subsection{IEEE-754}\label{ss.ieee754}

The IEEE-754-1985 Standard for Floating Point Arithmetic %
(ANSI/IEEE Std 754-1985) was introduced in 1985 \cite{IEEE754_1985} %
and revised in 2008 \cite{IEEE754_2008}. %
We focus here only on the binary encodings for standard input sizes %
provided by the standard. %
It is possible, in a non-standard way, to interpret the IEEE-754 system as a single %
floating point system derived from an input size $N \geq 1$. %
See Table~\ref{t.ieee754parameters}, which uses %
the convenient value of the {\em exponent field size} $s_N$. %
The formula %
\begin{equation}
s_N = \floor*{N^{0.611-N/3200}}
\end{equation}
happens to fit the IEEE-754 specification for the standard input sizes ($N = 16, N = 32, N = 64, N = 128$) %
although it is not a part of the standard. 
\begin{table}[htb]
\begin{center}
\begin{tabular}{l|l|l|c}
$N$ & $p_N$ & $b_N$ & name, (informal name) \\ \hline
8  & 5 & 3 & - \\
16 & 11 & 15 & binary16 (half precision) \\
32 & 23 & 127 & binary32 (single precision) \\
64 & 52 & 1023 & binary64 (double precision) \\
128 & 112 & 16383 & binary128 (quad precision) \\
$\,\,\raisebox{5pt}{\vdots}$ & $\,\,\raisebox{5pt}{\vdots}$ & $\,\,\raisebox{5pt}{\vdots}$ & \\
$N$ & $N - s_N$ & $2^{s_N - 1} - 1$ & 
\end{tabular}\\
\end{center}
\caption{parameters of IEEE-754 system.}\label{t.ieee754parameters}
\end{table}

From the input field we extract three unsigned integers $s, n, $ and $m$ as subfields %
from left to right. For fixed $N$, each subfield has a fixed size 1, $s_N$, and $N - s_N - 1$, respectively, 
as in the illustration: %
$$
\includegraphics[width=0.4\textwidth]{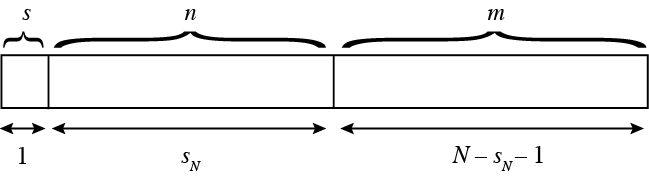}
$$
The value $x$ is then
\begin{equation}
x = \begin{cases}
(-1)^s \, (1 + m 2^{-N + s_N + 1}) \cdot 2^{n - b_N} & n \neq 0, n < 2^{s_N} - 1 \\
\pm0  & n = 0, m = 0 \\
(-1)^s \, m \cdot 2^{-N + s_N + 1 - b_N} & n = 0, m \neq 0 \\
\pm\infty & n = 2^{s_N} - 1, m = 0 \\
\text{NaN (not a number)} & n = 2^{s_N} - 1, m \neq 0
\end{cases}
\end{equation}

\begin{theorem}\label{t.ieeePBOM}
The precision by order of magnitude (PBOM) $B_N(n_x)$ %
for the IEEE-754 system %
is given by %
the rule
\begin{equation}
B_N (n_x) = \begin{cases}
0 & n_x \leq -N + s_N - b_N \\
N - s_N + b_N + n_x & -N + s_N - b_N < n_x \leq -b_N \\
N - s_N & -b_N < n_x < 2^{s_N} - 1 - b_N \\
0 & n_x \geq 2^{s_N} - 1 - b_N
\end{cases}
\end{equation}
\end{theorem}

\subsection{Posit System}{\label{ss.posit}

The Posit system %
is a component of Gustafson's comprehensive %
Unum program \cite{GustUnumBook, GustUnumRadical}.  %
It is designed as a one-to-one replacement for %
IEEE-754 floating point numbers. %
The Posit system forms a chapter in the recent revival %
of research into tapered number systems \cite{VariablePrecisionReport2019}. %
In the Posit system, %
the exponent field size parameter $s_N$ also appears, %
although in a different guise: now it only encodes the exponent %
to within a given regime of the dynamic range. %
In this setting, it is denoted $es$. %
The following formula for $es = es_N$ from \cite{PositArithmetic} %
is recommended for the Posit system \cite{PositStandardDocs3-2} for standard input sizes. %
\begin{equation}\label{e.posites}
es = es_N = \log_2 (N) - 3
\end{equation}
This gives the values $es$: 1, 2, 3, 4 for the standard input sizes %
$N = 16, N = 32, N = 64, N = 128$ %
and $es = 0$ for the ultra-low precision input size $N = 8$. %
Other choices for $es$ are possible, but we do not consider other $es$ values here. %
Roughly speaking, the choice of $es$ is a trade-off between precision and dynamic range. %

\begin{figure*}[thb]
\begin{center}
\includegraphics[width=0.99\textwidth]{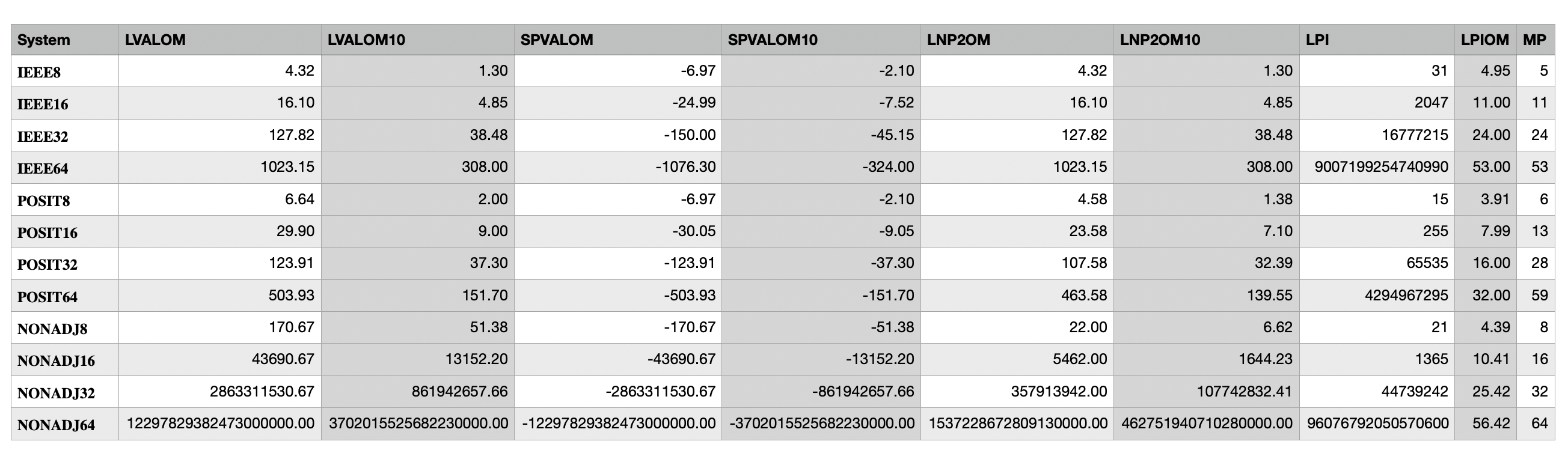}
\end{center}
\caption{Factors of merit comparison of IEEE, POSIT, and proposed NONADJ %
system for representative input sizes.}\label{f.properties}
\end{figure*}

A binary bit field of input size $N$ may be read from left to right as a posit number %
as follows. %
The first bit is a sign bit $s$. 
The next run following $s$ %
is interpreted as a redundant-signed \mbox{base-1} integer. %
That is, the length of the run after the sign bit's length %
presents a figure $r$, %
and the ``sign'' of the run (whether it is a run of ones, 1, %
or a run of zeros, 0) presents another figure $r_s$. %
After this, the bit which terminates the run (if any) %
is thrown away; we refer to this bit %
as the {\em stop bit} although this nomenclature is not part of the standard. %
It is possible that there is no stop bit. %
In this case, $r = N-1$. In that case if $r_s = 0$, 
it is exceptional, otherwise it is treated normally, %
which explains the one appearing in the second %
branch in the definition of $k$ below. %
(It also supplies the possibility $k = 0$.) %
From these, the figure $k$, ranging from $1 - N$ to $N-2$, is obtained:
\begin{equation}\label{e.positk}
k = (-1)^{r_s + 1} r + r_s = \begin{cases}
-r & r_s = 0 \\
r - 1 & r_s = 1
\end{cases}
\end{equation}
The $r_s$ value might be called the {\em positivity}, but this is not part of the standard. %
The remainder of the bits is read in one of two ways: 
if there is strictly more than $es$ bits, it is divided into two parts: %
the first part has length $es$ and the second part %
has length $N - r - es - 2$. %
The first part presents a figure $n$ as an unsigned integer, %
and the second part presents a figure $m$, also an unsigned integer. %

If there are exactly $es$ bits or fewer than $es$ bits %
(in this case the value is in the extremities of the dynamic range), %
all of the bits are used to present a figure $n$ as an unsigned integer %
of length $es$, %
but if there are fewer than $es$ bits available, %
the figure $n$ is obtained by right-padding with zeros, as needed. %
Finally, $m$ may be taken to be zero. %

Thus the encoding presents the figures $s, r, r_s, n, $ and $m$, %
as in the illustration:
$$
\includegraphics[width=0.4\textwidth]{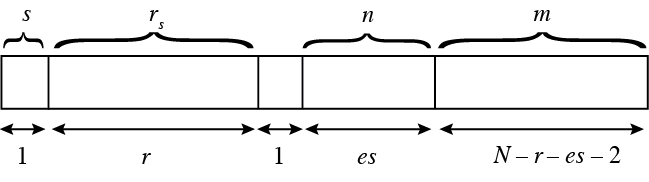}
$$
From this the value is
\begin{equation}\label{e.positx}
x = \begin{cases}
0 & 
\begin{tabular}[t]{l}
$k = 1 - N,$ \\ 
$s = 0$
\end{tabular}
\\
\text{NaR (not a real)} & 
\begin{tabular}[t]{l}
$k = 1 - N,$ \\ 
$s = 1$
\end{tabular}
\\
(-1)^s \, (1 + m \, 2^{-\log_2 (m) - 1}) \cdot 2^{k \cdot 2^{es} + n} & 
\text{       otherwise}
\end{cases}
\end{equation}
where we take $0 \cdot 2^{\infty} = 0$. %
We can see that the %
exponent in the Posit system %
may be broken into a regime %
whose size is determined by $es$ %
and which is indexed by $k$, %
while $n$ is an offset within the regime. %
There are always exactly $2N - 3$ regimes, 
but some regimes at the extremities are ``clipped'' %
in the sense that there is less than the full range of the offset value. %
It is not difficult to see from this description %
that the Posit system %
provides a unique code for $2^N - 1$ real numbers, %
and one exceptional code, NaR. %
It exhaustively uses up the possible %
bit fields for the given input size. %

Also note that due to the branch in the definition of $k$, %
there is a small asymmetry in the nominal precision between the case %
when $x < 0$ and when $x \geq 0$, when $1/x$ and $x$ are compared. %
However, because the sign of $n$ in Eq.~\ref{e.positx} is %
unaffected by this, the ranges where precisions are equal %
overlap only up to their endpoints. %
In other words, if $n_x$ is considered, %
precision is equal in two counterpart intervals, %
very nearly symmetric around $n_x = 0$, %
namely where $n_x$ is in the half-open interval $[\,k2^{es}, (k+1)2^{es})$ %
and the counterpart interval $[(-k-1)2^{es}, (-k)2^{es})$ %
where the regime is in fact not $-k$ but rather $-k-1$. %
For example, precision is at the maximum precision level %
in the regime $k=0$, corresponding to $n_x$ %
in the interval $[0, 2^{es})$. %
It is also at the maximum precision in its counterpart regime $k = -1$, %
corresponding to $n_x$ in the interval $[-2^{es}, 0)$, %
or in other words, in the half-open interval $[-2^{es}, 2^{es})$. %

We have the following. 

\begin{theorem}\label{t.positPBOM}
The precision by order of magnitude (PBOM) $B_N(n_x)$ for the system POSIT %
is given by %
the maximum between zero, and the rule
\begin{equation}\label{e.positPBOM}
B_N (n_x) = N + (-1)^{r_s} \, \floor*{n_x \cdot 2^{-es}} - r_s - es - 1
\end{equation}
where
\begin{equation}
r_s = \begin{cases} 1 & n_x \geq 0 \\ 0 & n_x < 0 \end{cases}
\end{equation}
\end{theorem}

{\em Proof:} See appendix~\ref{s.proof}. 

\subsection{The Nonadjacent System}\label{ss.nonadj}

We already introduced the nonadjacent system (NONADJ) in section~\ref{s.purereal}. %
We have the following re-expression of Theorem~\ref{t.precision} %
(some additional remarks are included in appendix~\ref{s.proof}): %

\begin{theorem}\label{t.nonadjPBOM}
The precision by order of magnitude (PBOM) $B_N(n_x)$ %
for the system NONADJ %
is given by the maximum between zero, and the rule
\begin{equation}\label{e.nonadjPBOM}
B_N (n_x) = N - \size(n_x)
\end{equation}
\end{theorem}

\subsection{Comparison of Systems}

In this section, we %
focus on the IEEE-754 system (IEEE), the Posit system (POSIT), and the newly proposed nonadjacent system, %
for which it suffices to consider the basic real nonadjacent system of section~\ref{s.purereal} (NONADJ). %

A fair comparison of the newly proposed nonadjacent system %
and the other two systems is a significant challenge, since the %
underlying technology cannot be treated as an invariant of the study. %
One system, the nonadjacent system, relies on ternary technology, %
while the others rely on binary technology. %
In lieu of efforts towards this end, %
we can consider here a handful of factors of merit that may be readily %
derived by analytic means from the definitions of the respective systems. %

The $B_N (n_x)$ curves for the three systems IEEE, POSIT, and NONADJ are shown %
for the ultra-low precision case $N=8$ in Fig.~\ref{f.BcurvesN8}. %
In Fig.~\ref{f.Bcurves}, %
the same curves are shown, along with the $B_N (n_x)$ curves for the standard input sizes $N=16, N=32,$ and $N=64$. %

We also consider a small number of factors of merit for each system, %
for four input sizes 8, 16, 32, 64. %
See Fig.~\ref{f.properties} for a concise summary %
of the values of the factors. %
The factors of merit that we consider for the three systems are: %
\begin{enumerate}
\item 
The largest value (LVAL). %
The largest finite real number that may be represented in the system. %
In Fig.~\ref{f.properties}, we include only the order of magnitude (the \mbox{base-2} logarithm), LVALOM. %
For convenience, we also show the \mbox{base-10} logarithm, LVALOM10. %
\item 
The smallest positive value (SPVAL). %
The smallest positive value that may be represented in the system. %
For IEEE-754, we can distinguish two cases, the smallest positive normal value, %
and the smallest positive subnormal value, while of course +0 is not considered; %
only the smallest subnormal value is shown in Fig.~\ref{f.properties}. %
SPVAL and LVAL together establish the total dynamic range. %
In Fig.~\ref{f.properties}, we include only the order of magnitude (the \mbox{base-2} logarithm), SPVALOM. %
It also presents the \mbox{base-10} logarithm, SPVALOM10. %
\item
largest non-power of two (LNP2). %
The largest finite representable value which is not a power of two. %
This figure assists in evaluating the tapered systems. %
For both the Posit system and the proposed nonadjacent system, %
it indicates a prominent marker past which the dynamic range is really only very sparsely populated. %
In particular, for the Posit system, it also %
indicates precisly where clipping effect begins to occur within the regimes due to right-padding. %
In Fig.~\ref{f.properties}, we show the order of magnitude (the \mbox{base-2} logarithm), LNP2OM. %
For convenience, we also show the \mbox{base-10} logarithm, LNP2OM10. %
\item 
largest precise integer (LPI). %
The largest integer that may be represented in the system, %
and whose immediate predecessor integer may also be represented in the system. %
This figure gives an idea of the range for integer arithmetic on the FPU %
in the given system. %
The LPI and the LPI's order of magnitude (the \mbox{base-2} logarithm), LPIOM, are presented %
in Fig.~\ref{f.properties}. %
\item
maximum precision (MP). %
The maximum precision, in binary or redundant signed binary digits, achievable %
in the system: precision in the system's ideal range. %
\end{enumerate}

\section{Conclusion}\label{s.conc}

We have proposed a new encoding %
for floating point numbers, %
based on the redundant signed radix~2 number system, 
using the canonical recoding, also known as the nonadjacent form. %
This system has long been known as a %
valuable resource for fast computer arithmetic, %
and our work has explored the possibility of its use outside of that context. %
The redundant nature of the system %
is what gives it both %
high precision and high dynamic range, %
because it allows for two detectable variable-width subfields in the input field. %
The resulting system %
implements round-ties-to-even rounding policy %
via truncation, %
making unbiased rounding both fast and easy to implement. %
The encoding offers an interoperable range %
of choices suitable for potential hardware targets, from embedded, to CPU, %
to scientific/accelerator/GPU devices. %
It achieves this via an extensible system %
that is easy to maintain, %
based on the proposed concept of point and point types. %
The point concept is closely analogous to a decimal point, %
and helps to establish an extensible tapered floating point system %
with properties facilitating error analysis, %
potentially offering aid and benefit to hardware designers, %
scientists and mathematicians, numerical application programmers, 
and system developers. %

%

The proposed system 
enjoys the stability and regularity properties of a tapered floating point system, %
including most notably gradual underflow and overflow. %
Unlike the binary encodings for tapered %
systems that were reviewed, and those known to the author, %
the present system encodes the exponent and significand %
in a directly recoverable way, offsetting the cost to hardware %
of finding the variably-located point. %
The exponenent and significand %
are encoded in a base higher than~1, %
as in the Morris system, but unlike in the Morris system, %
there are no bits wasted or reserved in any part of the input dynamic range. %
As a result, the system exhibits both a high dynamic range %
and generally high level of precision. %
In fact, as we noted, the system exceeds both the IEEE-754 %
and the Posit system in dynamic range %
is equal or better in precision throughout its entire dynamic range, %
and offers higher maximum precision in its ideal range. %
The encoding system %
shares with the Posit system the property that comparison %
is achieved by bitwise comparison, with the exception of %
the exceptional values $\pm\infty$. %

We also saw that the proposed encoding merges into one system %
all of the various encodings for the basic number and logic types %
which are of importance in practice, in a robust and type-safe way: %
boolean fields (including boolean true and boolean false %
which are distinguished from numerical 1 and numerical 0, %
and thereby potentially machine addresses), %
integers, and real values as floating point. %
We note that, for example, %
both the integer system and floating point system %
share a common zero, namely, the field of only zeros, %
which is the unique way of encoding the number zero. %
All other integers are either uniquely representable as integers (if very large), %
or else have exactly two representations in the system, %
one as floating point, and one as integer. %
All other floating point numbers have a single unique representation, 
with no exceptional corrective ranges like the subnormal range of IEEE-754. %
Analytic formulas are readily available for the encoding's precision and exponent, %
given any input real value $x$ and these do not contain any arbitrary constants. %
In addition, the range of representable integers %
in the proposed integer encoding mirrors closely but slightly exceeds the range %
of the two's complement system, when comparing equal input sizes. %

The proposed system, however, can also go beyond integers and real numbers, %
and provide support for complex numbers, two-vectors, and %
(if the hardware chooses) four-vectors and quaternions. %
In a straightforward and evident manner, %
it could be extended still further to support eight-vectors and octonions %
(for advanced scientific applications) %
while still maintaining perfect interoperability with lower-level encodings. %
The system's integration with these structures offers potential for intelligent and robust %
support for mathematical operations, %
including inverse operations like the square root and complex roots in general, %
and simple solvers, %
as well as elementary functions and numbers to a power (which can be complex, in general). %
This makes the full system a candidate for applications where robust, stable numerical processing %
is needed, including scientific computing, signal processing, %
computer graphics, and machine learning. %

For machine learning the system offers some particular advantages. %
First, during training regularization is used to keep weights %
bounded to a range near the values 1 and $-1$. %
As noted, this is precisely where the tapered precision is at its maximum, 
namely the full input size $N$. %
This offers the opportunity to save hardware area by decreasing the total size of stored values. %
If computation can be expected to remain in this range, %
or be in the range $[-1, 1]$ with some reasonable precision near zero, %
the computation can be carried out with a precision low enough to allow the hardware 
to be implemented in very low-area, low-energy environments, such as embedded/IoT applications, %
and for AI on the Edge. %
Further study of this question is warranted for such ultra-low precision applications. %



The use of nonadjacent forms thus offers computational advantages %
which offset the resource underutilization that nonadjacent %
nonzero bits will generate. %
Evaluation of this trade-off in terms of %
real physical space and time %
(area and delay, in the language of circuits) %
awaits technology on which such estimates could be based. %
It is an open question whether a ternary logic technology %
exists which makes the proposed system comparable %
by metrics such as area, energy/power, cost, and delay, with pure binary encodings. %
The ternary technology %
that the proposed system relies on can be implemented in a Field-Programmable Gate Array %
or other programmable hardware, %
but it requires doubling the input width size to account for binary-to-ternary encodings. %
To the best of our knowledge, %
true ternary technology is not currently supported by technology vendors 
or technology development platforms. %

\appendix

\section{Appendix A}\label{s.proof}


\begin{IEEEproof}[Proof of Theorem~\ref{t.wd}]
Let the width of an integer in nonadjacent form be $\width(x)$. %
For the proof we can assume that $x > 0$. %
When we add the values $J_n$ as $n$ grows, we have a geometric series, %
however, the term $(-1)^i$ separates into two cases depending on the parity of $n$. %
The following formulas result, which give the value $x$ of the maximal number (measured by value) in each batch of numbers having width $N$. %
In other words, it is the value of the last integer with the given width $N$. 
\begin{equation}\label{e.xlast}
x^{last}(N) = \begin{cases} \frac{2}{3}(2^N - 1) & $N$ \text{ even,} \\ \frac{1}{3} (2^{N+1} - 1) & $N$ \text{ odd.} \end{cases}
\end{equation}
If we add one to each of these, we obtain formulas for the minimal numbers (measured by value) in each batch of numbers of $N$ digits, 
for some $N$, and these are symmetric with the last numbers: 
\begin{equation}\label{e.xfirst}
x^{first}(N) = \begin{cases} \frac{1}{3}(2^{N + 1} + 1) & $N$ \text{ even,} \\ \frac{2}{3} (2^{N} + 1) & $N$ \text{ odd.} \end{cases}
\end{equation}
Now we claim that, for all integers $x$, 
\begin{equation}
\width(x) = \floor*{\log(\frac{3x-1}{2})}
\end{equation}
We can first invert the equations in \ref{e.xfirst} to give the (exact) width of the minimal number (by value) in each batch given its value $x$. Calling this value $W(x)$, we have two rules from the two expressions in equation \ref{e.xfirst}, namely, 
\begin{align}
W_1(x) &= \log(\frac{3x-1}{2})\\
W_2(x) &= \log(\frac{3x-2}{2})
\end{align}
By definition, the expression $W_2(x) = \log(\frac{3x-1}{2})$ %
returns an integer for every input $x$ which is a minimal number for some even number $N$ of digits. %
Therefore, for every such integer (where the number of digits in the system grows by 1), %
that value is the width $\width(x)$. The floor of that expression also provides the value $\width(x-1)$ for the maximum value %
of a batch of odd-sized digits, because $W_2(x)$ increases strictly. 

On the other hand, as the $x$ values transition from an even width batch to an odd-width batch, %
the value $W_1(x)$ passes an exact integer value, see Fig.~\ref{f.values}. %
Since $W_2(x-1) < W_1(x) < W_2(x)$, this proves the claim holds for all values of $x$. %
To establish formula (\ref{e.rel1}), observe that %
$$W_1(x) < \log(3x/2) < W_2(x+1)$$ %
Formula (\ref{e.rel2}) then follows immediately, %
since for integers $x$, the expression $\log(3x/2)$ is never an integer since $3$ divides $3x/2$.
\end{IEEEproof}

\begin{figure}[tb]
\begin{center}
\begin{tabular}{llll}
$x$ 		& $\log(\frac{3x-1}{2})$ 	& $\log(\frac{3x-2}{2})$ 	& $\log(\frac{3x}{2})$	\\ \hline
2 		& 1.321928			& 1					& 1.584962	\\
3		& 2					& 1.807354			& 2.169925	\\
4		& 2.459431			& 2.321928			& 2.584962	\\
5		& 2.807354			& 2.700439			& 2.906890	\\
6		& 3.087462			& 3					& 3.169925	\\
7		& 3.321928			& 3.247927			& 3.392317	\\
8		& 3.523561			& 3.459431			& 3.584962	\\
9		& 3.700439			& 3.643856			& 3.754887	\\
10		& 3.857980			& 3.807354			& 3.906890	\\
11		& 4					& 3.954196			& 4.044394	\\
12		& 4.129283			& 4.087462			& 4.169925	\\
13		& 4.247927			& 4.209453			& 4.285402
\end{tabular}
\end{center}
\caption{Values of $\log((3x-1)/2)$, $\log((3x-2)/2)$, $\log(3x/2)$ for $2 \leq x \leq 13$.}\label{f.values}
\end{figure}

\begin{IEEEproof}[Proof of Theorem~\ref{t.jacobsthal}]
The following proof appears in \cite{ShallitPrimer}. 
First observe from the initial part of the sequence of positive integers in nonadjacent form, 
$$1, 10, 10\UU, 100, 101, 10\UU0,\dots$$
that there is precisely one integer with width 0, and precisely one integer having width 1. %
Now let $x$ have width $n$. Suppose that $x$ ends in the digit $0$. In that case, the representation consists of a valid representation of width $n-1$ concatenated with $0$. Now suppose that it ends with a $1$ or a $\UU$. Then the representation of $x$ consists of a valid representation of width $n-2$ concatenated with either $01$ or $0\UU$. Therefore, if $J_n$ is the number of integers with width $n$ will satisfy 
$$J_n = J_{n-1} + 2 J_{n-2}$$
for $n > 1$. 
The resulting integer sequence has the closed form given in Eq.~\ref{e.jacobsthal}. 
\end{IEEEproof}

\begin{IEEEproof}[Proof of Theorem~\ref{t.positPBOM}]
For the system POSIT, %
the regime value $k$ is the value $n_x$ right-shifted. %
We can write
$$
k = \floor*{n_x \cdot 2^{-es}}
$$
and
$$
r_s = \begin{cases}
1 & n_x \geq 0 \\
0 & n_x < 0
\end{cases}
$$
Then
$$
r = (-1)^{r_s + 1} \, k + r_s
$$
and then from arithmetic on the bit field,
$$
p = N - r  - es - 2 + 1
$$
Collecting these together gives Eq.~\ref{e.positPBOM}, as desired. %
\end{IEEEproof}


\begin{IEEEproof}[Proof of Theorem~\ref{t.nonadjPBOM}]
For the system NONADJ with input size $N$, %
we have for $x > 0$ 
$$p(x) = N - \size(\width(x))$$
and $\width(x) = \width(2^{n_x})$. Have
\begin{align*}
p(n_x) 
&= N - \size(\floor*{\log_2 \frac{3}{2} \cdot 2^{n_x}}) \\
&= N - \size(\floor*{\log_2 (3) + n_x - 1}) \\
&= N - \size(n_x)
\end{align*}
So Eq.~\ref{e.nonadjPBOM} is obtained, %
following after the basic invariant equation $N = |n| + |m|$ in section~\ref{s.purereal}. %
\end{IEEEproof}

\bibliographystyle{IEEEtran}
\bibliography{mathmain}


\end{document}